\documentstyle[fleqn,11pt]{article}
\newcommand{\beq}{\begin{equation}}
\newcommand{\beqa}{\begin{eqnarray}}
\newcommand{\eeq}{\end{equation}}
\newcommand{\eeqa}{\end{eqnarray}}

\title{\bf Microlensing by brown dwarfs: the case of dark clusters}

\author{Alain Bouquet\thanks{Laboratoire de Physique Th\'eorique et
Hautes Energies, Universit\'es Paris 6 and Paris 7, Unit\'e associ\'ee au
CNRS (UA 280), 2 Place Jussieu, 75251 Paris Cedex 05 FRANCE} \and Jean
Kaplan$^*$ \and Jean-Manuel Verschaeve\thanks{Ecole Polytechnique, 91128
Palaiseau Cedex, FRANCE}}
\date{}
\begin{document}
\maketitle

\begin{abstract}
Experiments are looking for diffuse brown dwarfs in the dark galactic
halo through the gravitational lens effect. If brown dwarfs are clumped in
dark clusters, the event rate is not changed, but events are spatially
clustered, and stars nearby a micro-lensed one are likely to be
micro-lensed in a near future. Therefore, an intensive survey of the region
where a micro-lensing occurred should reveal many other events.
\end{abstract}
\noindent Abridged and updated version of {\bf ``M\'emoire de fin
d'\'etudes''}\\ by Jean-Manuel Verschaeve, Ecole Polytechnique,
Palaiseau, July 1991.\\
astro-ph/9402029

\vspace{0.5cm}

\section{Introduction}

The least exotic dark matter candidate is baryonic matter, under the form
of compact objects such as very massive black holes or brown dwarfs
(Carr, Bond \& Arnett 1984). Two teams reported the possible detection of
brown dwarfs in our galactic halo through gravitational micro-lensing
(Alcock et al. 1993, Aubourg et al. 1993): the light of a star is
amplified in a characteristic way when a brown dwarf crosses its line of
sight (Paczy\'nski 1986). Brown dwarfs could be clumped in dark clusters
similar to globular clusters (Carr \& Lacey 1987) and we explore the
observational consequences of such dark clusters. We conclude that:\\
i) the event rate is not changed.\\
ii) micro-lensing events should cluster on a few spots on the sky.\\
iii) stars which are near a lensed star are likely to be lensed in
a near future by a brown dwarf of the same cluster.

\section{Event rate}

The gravitational lens effect is a consequence of the deflection of light
rays by massive objects. When a brown dwarf comes at a distance $R$ to
the line of sight to a star, the light from this star is amplified by a
factor $A(R)$:
\beq
A(R) = \frac {R^2+2R_{\rm E}^2}{R\sqrt{R^2 + 4R_{\rm E}^2}}
\eeq
where the Einstein radius $R_{\rm E}$ is defined by:
\beq
R_{\rm E}^2 = \frac{4 G M_{\rm bd}}{c^2} \frac{D_{\rm bd}
 (D_{\rm star}-D_{\rm bd})}{D_{\rm star}}
\eeq
$M_{\rm bd}$ is the mass of the brown dwarf, $D_{\rm bd}$ its distance
from the observer, and $D_{\rm star}$ the distance of the star. The
dimensionless impact parameter $u$ is defined as:
\beq
u(A) = \frac{R}{R_{\rm E}} = \left[ \frac{2 \, A}{\sqrt{A^2-1}} - 2 \,
\right]^{1/2}
 \label{uA}
\eeq
A micro-lensing event is detected if the amplification is larger than
some threshold $A$. The event rate is the number $\Gamma(A)$ of
detectable events per star and per unit time, and is given by the number
of brown dwarfs which enter per unit time the micro-lensing ``tube'' of
radius $R(A,D_{\rm bd}) = u(A) R_{\rm E}(D_{\rm bd})$ around the line of
sight to the star (Griest 1991):
\beq
\Gamma(A) = \int  2 \, u(A) \, R_{\rm E}(D_{\rm bd}) \, V_{\perp} \;
n_{\rm bd}(D_{\rm bd}) \; {\rm d}D_{\rm bd}
\label{gamma}
\eeq
where $V_{\perp}$ is the transverse velocity of brown dwarfs, and $n_{\rm
bd}(D_{\rm bd})$ is their number density at distance $D_{\rm bd}$. The
number density $n_{\rm bd}$ is the same whether
brown dwarfs are clustered or not. Therefore the event rate does not
depend on the clustering.

We assume that the mass density profile of the halo is well described by
an approximate isothermal distribution with a core radius $a= 5 \pm 3$
kpc (Bahcall \& Soneira 1980, Caldwell \& Ostriker 1981):
\beq
\rho(r) = \rho_\odot \frac{r_\odot^2+a^2}{r^2+a^2}
\label{rho}
\eeq
The halo mass density in the solar neighbourhood is estimated to be
$\rho_\odot = 0.008 \, M_\odot/{\rm pc}^3$ (Flores 1988), and the
distance between the Sun and the galactic center is $r_\odot$ = 8.5 kpc.
The possible mass $M_{\rm bd}$ of a brown dwarf ranges from 10$^{-7}$
$M_\odot$ (evaporation limit, De R\'ujula et al. 1992) to 10$^{-1}$
$M_\odot$ (hydrogen burning limit). Then:
\beq
\Gamma(A) = \Gamma_0 \, u(A) \, \frac{V_{\perp}}{200 \, {\rm km/s}}\left[
\frac{0.1 \, M_\odot}{M_{\rm bd}} \right]^{1/2}
\label{gammanum}
\eeq
where $\Gamma_0 = 16 \times 10^{-14} \, {\rm s}^{-1}$ for stars in the
LMC. Table 1 shows the variations of $\Gamma_0$ with
the core radius $a$, for stars in the LMC and in the galactic center.
We assumed that all brown dwarfs had the same transverse velocity
$V_{\perp} = 200$ km/s. It is possible to include the actual velocity
distribution of the halo and to take into account the velocities of the
star and of the Earth  (Griest 1991), but this is an unnecessary
complication at this level. The approximation of identical velocities is
better for brown dwarfs inside a cluster than for diffuse ones, because
the velocity dispersion $V_{\rm bd}$ inside a cluster of radius $R_{\rm
tidal}$ is small compared to the transverse velocity $V_{\perp} \simeq
200 \, {\rm km/s}$ of the cluster:
\beq
V_{\rm bd} < \left[ \frac{G M_{\rm cluster}}{R_{\rm tidal}} \right]^{1/2}
\simeq 21 {\rm km/s}\left[ \frac{M_{\rm cluster}}{10^6 \, M_\odot} \,
\frac{10 {\rm pc}}{R_{\rm tidal}} \right]^{1/2}
\eeq

The expected number of events for $N_{\rm stars}$ stars monitored during a
time $t_{\rm obs}$ is:
\beqa
N_{\rm events} &=& \Gamma(A) \; N_{\rm stars} \; t_{\rm obs}\\
 &\simeq& 13 \; u(A)  \, \left[ \frac{0.1 \, M_\odot}{M_{\rm bd}}
\right]^{1/2} \; \frac{N_{\rm stars}}{5\times10^6 \, {\rm stars}} \;
\frac{t_{\rm obs}}{180 \, {\rm nights}} \,(\times \,{\rm Detection \,
efficiency})
\label{Nevent}
\eeqa
The EROS team uses a CCD camera with a field of view of about 0.4 square
degree to monitor about $8\times 10^4$ stars every half hour in the LMC
bar, to be sensitive to lighter brown dwarfs ($M_{\rm bd} = 10^{-6 \pm 1}
\, M_\odot$), and a Schmidt telescope with a field of view of 25 square
degrees to monitor $5 \times 10^6$ stars every night to be sensitive to
heavy brown dwarfs ($M_{\rm bd} = 10^{-2 \pm 1} \, M_\odot$). The MACHO
team gets the same number of stars using a CCD camera to covers about 60
fields (0.5 square degree each) once per night.

\section{Space clustering}

Lensing events will obviously be more concentrated on a few spots on the
sky if brown dwarfs are clustered than if they are diffuse. How many
clusters lie in the area surveyed by experiments? The mean number $N_{\rm
clusters}$ of dark clusters in a narrow cone of solid angle d$\Omega$ is
just the integral along the line of sight of the number density of dark
clusters:
\beqa
N_{\rm clusters} &=& \int \frac{\rho(D)}{M_{\rm cluster}} \, D^2 \; {\rm
d}D \, {\rm d}\Omega \\
 &=& 9.3 \,{\rm clusters} \; \frac{10^6 \, M_\odot}{M_{\rm cluster}} \,
\frac{{\rm d}\Omega}{1 \,{\rm square \,degree}}
\label{nclusters}
\eeqa

Searches for heavy brown dwarfs scan an area of several square degrees
where about 230 dark clusters of 10$^6$ $M_\odot$ lie, and more if they
are lighter. It is not surprising that the few claimed events (Alcock et
al. 1993, Aubourg et al. 1993) do not appear to be concentrated. The
eventual presence of dark clusters will not be apparent before statistics
are increased by one order of magnitude. On the other hand, light brown
dwarf searches scan a smaller area of 0.4 square degree, where 3 to 5
clusters of 10$^6$ $M_\odot$ are expected. There might even be no cluster
in the field, although the probability of such a disaster is low (between
5\% and 1\%, from Poisson statistics). The good point of having few
clusters is that even a small number of events will be very clustered on
the sky. Such a pattern will be a clear signature of the clustering of
brown dwarfs. Let us stress that the probability to find at least 2
events at the same spot is not negligible, even for few events and many
clusters: there is a 42\% probability for instance that at least 2 events
out of 5 are at the same spot if there are 20 clusters in the field (see
Table 1).

These results depend weakly on the halo core radius $a$ and on the target
direction (see Table 2) but they depend strongly on the mass of dark
clusters. Since we know nothing about eventual dark clusters, we just
assume that they are similar to globular clusters (Harris \& Racine 1979,
Djorgovski 1988). These have masses $M_{\rm cluster}$ ranging between
10$^4$ and 10$^6$ $M_\odot$, an external (tidal) radius $R_{\rm tidal}
\simeq 10$ pc, and a core radius $R_{\rm core} \simeq 0.5$ pc (the core
radius is defined as the radius where the surface density drops to half
its central value (King 1966), and more massive clusters tend to have
smaller cores). The core contains nearly half the cluster mass. The
angular diameter of such clusters is small:
\beqa
\theta_{\rm core} &=& 20\,{\rm arcsec} \;\; \frac{R_{\rm core}}{0.5 \,
{\rm pc}} \, \frac{10 \, {\rm kpc}}{D_{\rm cluster}}\\
\theta_{\rm tidal} &=& 400\,{\rm arcsec} \;\; \frac{R_{\rm tidal}}{10 \,
{\rm pc}} \, \frac{10 \, {\rm kpc}}{D_{\rm cluster}}
\eeqa
Clusters do not overlap on the sky: the mean angular distance between 2
clusters is 20 arcmin (= $1/\sqrt{N_{\rm clusters}}$ degree), whereas the
angular diameter of the cluster core is 20 arcsec. The mean angular
separation of monitored stars is about 8 arcsec ($2\times 10^5$ stars
are monitored per square degree in the LMC). Therefore, there is a
high probability that at least one monitored star lies behind each dark
cluster core.

If $N_{\rm events}$ lensing events are observed, the mean angular
separation between them is expected to be:
\beq
\theta_{\rm diffuse} = \left[ \frac{{\rm Total \, area}}{N_{\rm events}}
\right]^{1/2} \simeq 95  \, {\rm arcmin} \,  \left[ \frac{10}{N_{\rm
events}} \right]^{1/2}
\eeq
for diffuse events, and:
\beq
\theta_{\rm clustered} = \frac{\theta_{\rm core}}{ \left[ N_{\rm
events}/N_{\rm clusters} \right]^{1/2} }  \simeq 1.6 \, {\rm arcmin} \,
\left[ \frac{10}{N_{\rm events}} \right]^{1/2}
\eeq
for events due to the {\em same} dark cluster. Note that the ratio of
these angles:
\beq
\frac{\theta_{\rm diffuse}}{\theta_{\rm clustered}} = \frac{1}{\theta_{\rm
core} \sqrt{{\rm d}N_{\rm clusters}/{\rm d}\Omega}} \simeq 60
\eeq
is quite large and independent of the number $N_{\rm events}$ of events
and of the size ${\rm d}\Omega$ of the surveyed area.

\section{Time clustering}

Once a micro-lensing event is detected, the strategy should change
if we assume that it is due to a dark cluster. This cluster is at a fixed
distance from us, and there is a negligible probability that another
cluster overlaps on the same line of sight, as we just saw. The rate of
subsequent events in the same direction is then different, because we no
longer integrate over the line of sight. We now expect more events near
the cluster, and less events away from it. Two kinds of repetitive
events can be expected:\\
\indent i) the lensing of a nearby star by the same brown dwarf.\\
\indent ii) the lensing of a nearby star by another brown dwarf.

The first case happens whether brown dwarfs are clustered of not, and the
minimum time $t_{\rm wait}$ to wait between successive lensings is:
\beq
t_{\rm wait} = \frac{\theta_{\rm star}}{V_\perp/D_{\rm bd}} = 2000 \,
{\rm years} \frac{\theta_{\rm star}}{8 \, {\rm arcsec}} \, \frac{D_{\rm
bd}}{10 \, {\rm kpc}} \, \frac{200 \, {\rm km/s}}{V_\perp}
\eeq
where $\theta_{\rm star}$ is the angular distance between monitored
stars. This is a minimum time since there is no reason that the nearest
monitored star lie in the direction of the brown dwarf movement. Note that
this time is about the same if brown dwarfs belong to a thick disk
instead of a halo (the decrease in $D_{\rm bd}$ being compensated by the
decrease in $V_\perp$).

The second case, the lensing of a nearby star by another brown
dwarf, is specific of dark clusters.
The mean time $t_{\rm wait}$ to wait between two micro-lensings of the
same star (or between the micro-lensings of two {\em given}
stars) in the diffuse case is $\Gamma^{-1}$ by definition. Hence:
\beq
t_{\rm wait} = 200\,000 \, {\rm years}  \, \frac{1}{u(A)} \, \frac{200
\, {\rm km/s}}{V_{\perp}} \left[ \frac{M_{\rm bd}}{0.1 \, M_\odot}
\right]^{1/2}
\eeq
from Equation \ref{gammanum}.
If brown dwarfs are clustered, the event rate now is the rate at which
brown dwarfs enter in the area of radius $u(A) \, R_{\rm E}$ around the
star times the surface density $\Sigma$ of brown dwarfs in the cluster:
\beq
\Gamma = 2 \, u(A) \, R_{\rm E} \, V_{\perp} \, \Sigma = 2 \, u(A) \,
R_{\rm E} \, V_{\perp} \frac{M_{\rm cluster}}{M_{\rm bd}} \, \frac{1}{\pi
R_{\rm core}^2}
\eeq
The time $t_{\rm wait}$ between lensings is then much shorter:
\beqa
t_{\rm wait} &=& \frac{1}{\Gamma} = \frac{\pi \, c \, R_{\rm core}^2}{4
\, u(A) \, V_{\perp} \, M_{\rm cluster}} \left[ \frac{M_{\rm bd} \,
D_{\rm star}}{G \, D_{\rm bd} \, (D_{\rm star}-D_{\rm bd})}
\right]^{1/2}\\
 &\simeq& 32 \, {\rm years} \, \frac{1}{u(A)} \left[\frac{M_{\rm bd}}{0.1
\, M_\odot} \right]^{1/2} \frac{10^6 \, M_\odot}{M_{\rm cluster}}
\left[\frac{10 \, {\rm kpc}}{D_{\rm bd}} \frac{D_{\rm star}}{D_{\rm
star}-D_{\rm bd}} \right]^{1/2} \left[\frac{R_{\rm core}}{0.5 \, {\rm pc}}
\right]^2
\eeqa
and it depends sensitively on the cluster parameters $D_{\rm bd}$,
$R_{\rm core}$ and $M_{\rm cluster}$, and on the brown dwarf parameters
$V_{\perp}$ and $M_{\rm bd}$. An essential information on the latter will
be given by the duration $t_{\rm event}$ of the lensing event, which is
the mean time elapsed from detection (treshhold amplification $A$) to peak
amplification and back to amplification $A$ again. From simple geometry,
it is:
\beq
t_{\rm event} = \frac{\pi}{2} \, \frac{u(A) \, R_E}{V_{\perp}}
\label{tevent}
\eeq
for a brown dwarf of relative transverse velocity $V_{\perp}$ (typically
about 200 km/s). For a star in the Large Magellanic Cloud (LMC) at
$D_{\rm star} \simeq 50$ kpc, a lensing event lasts between a few hours
and a few weeks, depending on the minimal amplification $A$, and on the
mass and distance of the brown dwarf:
\beq t_{\rm event} \simeq 35 \,{\rm days} \; u(A) \; \frac{200 \, {\rm
km/s}}{V_{\perp}} \left[ \frac{M_{\rm bd}}{0.1 \, M_\odot} \frac{D_{\rm
bd}}{10 \, {\rm kpc}} \frac{D_{\rm star}-D_{\rm bd}}{D_{\rm star}}
\right]^{1/2}
\label{teventnum}
\eeq

The ratio $t_{\rm wait}/t_{\rm event}$:
\beq
\frac{t_{\rm wait}}{t_{\rm event}} = \frac{1}{\pi \, u^2 \, R_{\rm E}^2 \,
\Sigma} = 320 \, \frac{10 \, {\rm kpc}}{D_{\rm bd}} \, \frac{D_{\rm
star}}{D_{\rm star} - D_{\rm bd}} \, \frac{10^6 \, M_\odot}{M_{\rm
cluster}} \, \left[\frac{R_{\rm core}}{0.5 \, {\rm pc}} \right]^2
\eeq
is independent of the brown dwarf mass and velocity, but depends
strongly on the core radius of the cluster. Thus, in principle, if one
is lucky enough to observe a micro-lensing lasting for one day (as
expected from a 10$^{-4}$ $M_\odot$ brown dwarf), the same star (or any
nearby one) might well be micro-lensed every year for 5000 years
(the time needed to travel 1 pc at 200 km/s). It is therefore worthwhile
to follow a star the light of which has been amplified (even if there is
some doubt that the event was actually a micro-lensing), because another
micro-lensing would be very unlikely in the case of a diffuse
distribution of brown dwarfs. A star which undergoes such successive
``flashes'' cannot be confused with a genuinely variable star (like a
cepheid), because the time delay between flashes is irregular, and long
compared to the flash duration. It cannot be confused with a flare star,
because the amplification is time symmetric and achromatic.

Is there a risk that a star, already lensed by one brown dwarf, be
simultaneously lensed by another one? A double lensing is detected when a
brown dwarf gets within $u(A) \,R_{\rm E}$ distance to the line of sight
to the star, and another one is already inside. If brown dwarfs are not
correlated within the cluster (e.g. no binaries), the probability $P_{\rm
dl}$ of such an event is just the area of the disk of radius $u(A) \,
R_{\rm E}$, times the surface density $\Sigma$:
\beqa
P_{\rm dl} &=& \pi \, u^2 \, R_{\rm E}^2 \, \Sigma \\
 &=& \pi \, u^2 \frac{4\,G \, M_{\rm bd}}{c^2} \frac{D_{\rm bd} (D_{\rm
star}-D_{\rm bd})}{D_{\rm star}} \, \frac{M_{\rm cluster}}{M_{\rm bd}}
\frac{1}{\pi \,R_{\rm core}^2} \\
 &=& 3.1\times 10^{-3} \; u^2 \; \frac{D_{\rm bd}}{10 \, {\rm
kpc}}\frac{D_{\rm star}-D_{\rm bd}}{D_{\rm star}} \,\frac{M_{\rm
cluster}}{10^6 \, M_\odot} \left[ \frac{0.5 \, {\rm pc}}{R_{\rm core}}
\right]^2
\eeqa
This probability is less than 1 \% . This is most welcome, since otherwise
the light curve would be asymmetrical, and rejected as noise by the
scanning algorithms of the brown dwarf experiments. Note that this
probability does not depend on the mass of the brown dwarf, but only the
the cluster parameters (mass, distance and radius).

\section{Unresolved stars}

The number of monitored stars behind a cluster
is small because the angular diameter of the core (20 arcsec) is
similar to the angular separation between monitored stars (8
arcsec). The density of brown dwarfs is large near the micro-lensed star,
which increases the probability of lensing nearby stars, but we
cannot use this advantage if there is no other monitored star nearby! One
could monitor more stars in the same area by an increase of the
sensitivity of the camera, but the stars then start to overlap and the
confusion limit is quickly reached when their distance becomes less than
the seeing (about one arcsec).

We suggested (Baillon et al. 1993) a more efficient way to detect
micro-lensings. In a crowded field of unresolved stars (e.g. the
Andromeda galaxy M31 or the bar of the LMC), one could detect the
luminosity increase on one pixel of the image due to the micro-lensing of
one of the unresolved stars on this pixel. There are typically about 10
stars on each pixel of a CCD picture of the LMC, but the amplification
required to detect faint ones is so large that the lensing probability
becomes negligible ($\Gamma(A) \propto 1/A$ when $A \gg 1$). It turns out
that about one star only per pixel can be sufficiently amplified to
become detectable.
Monitoring pixels instead of stars has several advantages if brown dwarfs
are clustered:\\
i) the number of pixels is much larger than the number of monitored
stars, hence a gain in detection efficiency.\\
ii) CCD pixels are contiguous, and there is one of them in front of any
cluster.\\
iii) CCD pixels are often less than 1 arcsec wide, smaller than the
angular size of a dark cluster which is then covered by several pixels.
As a by-product, such an experiment would then allow a study of (dark)
cluster cores without being hampered by the usual seeing problem.

\vspace{0.5cm}
{\bf Note added:} Eyal Maoz recently presented similar but more
optimistic conclusions because he takes the number of monitored stars to
be 10$^6$ per square degree (Maoz 1994). He also remarks that the
durations of micro-lensing events due to the same dark cluster should be
strongly correlated, a point that we overlooked.

\vspace{1cm}
{\Large {\bf References}}

\noindent Alcock C. et al {\em Nature} 365, 621.

\noindent Aubourg E. et al {\em Nature} 365, 623.

\noindent Ardeberg A. et~al., 1985, {\em Astronomy and Astrophysics} 148,
263

\noindent Bahcall J.N. and Soneira R.M., 1980,{\em Astrophysical Journal
Supp.} 44, 73

\noindent Baillon P. et al.,1993, {\em Astronomy and Astrophysics} 277, 1

\noindent Caldwell J.A.R. and Ostriker J.P., 1981, {\em Astrophysical
Journal} 251, 61

\noindent Carr B.J., Bond J.R., and Arnett W.D., 1984, {\em Astrophysical
Journal} 277, 445

\noindent Carr B.J. and Lacey C.G., 1987, {\em Astrophysical Journal}
316, 23

\noindent De R\'ujula A., Jetzer Ph. and Mass\'o \'E., {\em Astronomy and
Astrophysics} 254, 99

\noindent Djorgovski S ., ``Surface photometry of globular clusters'',
Proceedings of the Harlow Shapley Symposium on Globular Cluster Systems
(J.E.Grindlay and A.G. Davis Philip eds., I.A.U. 1988)

\noindent Fall S.M. and Rees M.J., ``The origin of globular clusters'',
Proceedings of the Harlow Shapley Symposium on Globular Cluster Systems
(J.E.Grindlay and A.G. Davis Philip eds., I.A.U. 1988)

\noindent Flores R.~A., 1988, {\em Physics Letters B} 215, 73

\noindent Griest K., 1991, {\em Astrophysical Journal} 366, 412

\noindent Hardy E. et~al., 1984, {\em Astrophysical Journal} 278, 592

\noindent Harris and Racine , 1979, {\em Annual Review of Astronomy and
Astrophysics} 17, 241

\noindent Hegyi D. and Olive K.A., 1986, {\em Astrophysical Journal}
303,56

\noindent King I.R., 1966, {\em Astrophysical Journal} 71,64

\noindent Kormandy J. and Knapp G.~R. (eds.), 1987, Proc. IAU Symp. 117,
Dark matter in the Universe. Reidel.

\noindent Maoz E., 1994, preprint astro-ph/9402027

\noindent Paczy\'nski B., 1986., {\em Astrophysical Journal} 304, 1.

\vspace{1cm}

{\Large {\bf Tables}}
\vspace{0.5cm}
\begin{center}
\begin{tabular}{|l|rrrr|}
 \hline
Number of clusters & 3	& 10	& 20	& 100\\
Number of events &	& & & \\ \hline
3	& 78\%	& 28\%	& 14\%	& 3\% \\
5	& 100\%	& 70\%	& 42\%	& 10\% \\
10	& 100\%	& 99.96\%	& 93\%	& 37\% \\
20	& 100\%	& 100\%	& 100\%	& 73\%	\\ \hline
\end{tabular}

{\bf Table 1:} Probability to find {\em at least} two micro-lensing events
at the same spot (within less than 1 arcmin) as a function of
the number of clusters and of the number of events.
\end{center}

\vspace{0.5cm}
\begin{center}
\begin{tabular}{|l||rrrr|}
\hline
Halo core radius $a$ (kpc)	& 0	& 2 &	5	& 8 \\ \hline \hline
$\Gamma_0$ ($\times10^{14}$)	in the LMC & 14 & 15 &	16 & 19
\\ \hline
$\Gamma_0$ ($\times10^{14}$)	in the Galactic Center &
$\infty$ & 24 & 10 & 7 \\ \hline \hline
Number of 10$^6$ $M_\odot$ clusters/square degree towards LMC	& 7.2	&
7.6	& 9.3	& 12.1 \\ \hline
\end{tabular}
\end{center}

{\bf Table 2:} Event rate $\Gamma_0$ per star as a function of the halo
core radius $a$, for stars in the LMC (at $D_{\rm star}$ = 50 kpc) and in
the Galactic Center (at 8.5 kpc). Also shown is the number of 10$^6 \,
M_\odot$ clusters per square degree in the direction of the LMC.

\end{document}